\documentclass[12pt]{article}

\usepackage{amssymb, amsmath}
\usepackage{authblk}
\newcommand{\beq}{\begin{equation}}  
\newcommand{\eeq}{\end{equation}}
\newcommand{\non}{\nonumber} 
\newcommand{\lmk}{\left(}  \newcommand{\rmk}{\right)}
\newcommand{\lkk}{\left[}  \newcommand{\rkk}{\right]}

\newcommand{\be}{\begin{eqnarray}}
\newcommand{\ee}{\end{eqnarray}}
\newcommand{\order}{{\cal O}}
\newcommand{\calR}{{\cal R}}
\newcommand{\calP}{{\cal P}}
\newcommand{\calN}{{\cal N}}

\newcommand{\mg}{M_{Pl}}
\newcommand{\mx}{M_{X}}
\newcommand{\calL}{{\cal L}}
\newcommand{\NS}{{\cal N}_S}
\newcommand{\NFL}{{\cal N}_{FL}}

\newcommand{\GFM}{\Gamma_{FM}}
\newcommand{\GSK}{\Gamma_{SK}}
\newcommand{\GSM}{\Gamma_{SM}}
\newcommand{\GSKM}{\Gamma_{SKM}}
\newcommand{\phis}{\phi_*}
\newcommand{\phif}{\phi_f}
\newcommand{\zetam}{\zeta_m}
\newcommand{\zetar}{\zeta_{R2}}
\newcommand{\nus}{\nu_*}
\newcommand{\fnl}{f_{NL}}
\newcommand{\fnlm}{f_{NLm}}
\newcommand{\nsr}{n_{sR2}}
\newcommand{\nsm}{n_{sm}}
\def\non{\nonumber}
\def\Slash#1{\hskip 0.05 cm \slash\hskip -0.25 cm #1} 

\begin{document}
\author{Yuki Watanabe$^1$ and Jun'ichi Yokoyama$^{1,2}$}
\affil{\it\small $^1$Research Center for the Early Universe (RESCEU),\\
\it\small Graduate School of Science, The University of
Tokyo, Tokyo 113-0033, Japan}
\affil{\it $^2$Kavli Institute for the Physics and Mathematics of the Universe
(WPI),\\
\it\small The University of Tokyo, Kashiwa, Chiba 277-8568, Japan}

\title{\bf Gravitational modulated reheating and non-Gaussianity 
in supergravity $R^2$ inflation}

\maketitle

\abstract{
Reheating after $R^2$ inflation proceeds through gravitational particle production of 
conformally noninvariant fields.  We argue that the nonvanishing expectation
value of flat directions generic in supersymmetric theories break
conformal invariance of the fields coupled to them in a position-dependent
manner due to  quantum fluctuations.  As a result modulated reheating
can occur after the supergravity $R^2$-inflation.  The resultant
curvature fluctuation is a mixture of the one produced during inflation
and that produced by modulated reheating.  The spectral index takes
a value between $n_s=0.960$ and 0.983, the nonlinearity parameter of the
local-type non-Gaussianity can be $\fnl\sim \pm 10$, and the
tensor-to-scalar ratio is $r \le 4\times 10^{-3}$.}

\begin{flushright}
RESCEU-5/13
\end{flushright}

\newpage
\section{Introduction}
The $R^2$ inflation \cite{Starobinsky:1980te} is a unique inflation
model in the sense that it is one of the oldest models of inflation
\cite{Guth:1980zm} but still observationally viable. In fact, it is not
only viable but preferred by the  observational results of 
WMAP9 \cite{Hinshaw:2012fq},\footnote{The latest observational results of Planck \cite{Ade:2013rta, Ade:2013ydc} also support the $R^2$ inflation, with tighter constraints on $n_s$, $r$ and $f_{NL}$. We discuss their results further in subsection~\ref{subsec:planck} .} with its predictions of scalar spectral
index,\footnote{The first prediction of the scalar spectral index for $R^2$ inflation was made correctly by Mukhanov and Chibisov \cite{Mukhanov:1981xt}} $n_s=0.964$, and the tensor-to-scalar ratio, $r=3.9\times
10^{-3}$, occupying 
the center of the likelihood contour. 
In order to further probe or falsify this model, it is desired 
to have more predictions that can be tested soon. 
Among such quantities is the nonlinearity parameter $f_{NL}$ of
curvature fluctuations \cite{Komatsu:2001rj} which measures
 the deviation of their statistical distribution from Gaussian.

Previously, $R^2$ inflation has been known to yield highly Gaussian
curvature fluctuations, 
which can be easily understood regarding the extra scalar degree of
freedom
in $f(R)$-type theory, dubbed {\em scalaron} $\varphi$, as the driving
force
of inflation \cite{Maeda:1988ab}.  
Indeed $\varphi$ has a very flat effective potential
ideal for chaotic inflation   \cite{Linde:1983gd} with a canonical
kinetic term.
In this paper we argue that nontrivial processes in the reheating phase
 after inflation may produce additional curvature perturbations, which
 can be appreciably
 non-Gaussian, through the so-called modulated reheating scenario 
\cite{Dvali:2003em} known to produce $f_{NL}$ of local type 
\cite{Zaldarriaga:2003my, Suyama:2007bg}.

In this theory, inflaton is followed by damped oscillation of  the
scalar 
curvature $R(t)$, which induces gravitational particle production of 
conformally non-invariant fields thereby reheating the Universe.
This process is equivalently described by the decay of the 
scalaron-inflaton $\varphi$ which oscillates around the potential
minimum after inflation. 
In the scalaron picture,
 $\varphi$ is coupled with conformally non-invariant fields only, so it is coupled with neither massless fermions nor gauge bosons (except gauge conformal anomaly \cite{Watanabe:2010vy}). 
Therefore,  the previous study of reheating after
$R^2$ inflation \cite{Vilenkin:1985md,Mijic:1986iv,Arbuzova:2011fu} including more general
$f(R)$ inflation \cite{Motohashi:2012tt} mostly
 considered creation of massless minimally coupled scalar bosons.

However, since mass terms break the conformal invariance, scalarons can decay into massive particles through their mass terms \cite{Gorbunov:2010bn} (see also \cite{Watanabe:2006ku, Faulkner:2006ub} for earlier works).
These channels open up a new possibility to produce curvature perturbations because the mass terms, which are determined by the Higgs field, turn out to be spatially dependent reflecting quantum fluctuations of the Higgs condensation \cite{Kunimitsu:2012xx} generated during inflation. 
As a result the modulated reheating \cite{Dvali:2003em} may take
 place which  contributes to the observed curvature perturbations.

Unfortunately, however, such a mechanism does 
not work in the original $R^2$ inflation model which is realized by adding a single term, $R^2/(6M^2)$, to the gravitational Lagrangian. In this simplest model the scalaron mass is determined (or bounded from above, in case there is other sources of curvature perturbation besides the scalaron's fluctuations) by the observed amplitude of curvature perturbations as $M \simeq 3\times 10^{13}$ GeV.
As a result, its decay rate per mode 
$\Gamma_1 = M^3/(192\pi M_{\rm Pl}^2) \simeq 8$ GeV is so small 
that the Universe would not be reheated until long after inflation. 
During this long period of scalaron oscillation, the Higgs condensation 
also oscillates and decreases its amplitude.
Therefore, by the time the reheating occurs, the standard model (SM) 
Higgs field value would be too small to leave any observational trace in
contrast to the case studied in \cite{DeSimone:2012gq}.
So much is the story of $R^2$ inflation in the SM.

On the other hand, it is barely likely that the SM is the ultimate theory with which we should describe the birth and evolution of the Universe.
As the primary extension of the SM, there are a number of sound motivations to introduce the supersymmetry (SUSY).
Then the $R^2$ inflation -- or its analogue -- should be realized in the context of local supersymmetry or the supergravity (SUGRA).
An attempt to realize $R^2$ inflation in this context
was first made by Ketov \cite{Ketov:2010eg}, and was improved
 by Ketov and Starobinsky \cite{Ketov:2010qz}.
Its most wonderful feature is  that the scalaron mass in the
reheating phase is not directly related to the amplitude of curvature
fluctuations and can be much heavier than  in the original model, so that the Universe can be reheated immediately after inflation by gravitational particle production.
The other plausible feature is the existence of flat-direction scalar fields which have potentials much flatter than the SM Higgs field has, even after the effects of SUSY breaking and non-renormalizable superpotential are taken into account.  
Thus quantum fluctuations around a large expectation value of the 
 flat direction can induce modulated reheating \cite{Enqvist:2003uk}, 
and generation of appreciable amount of local-type non-Gaussianity is 
possible in this model.

The rest of the paper is organized as follows. 
In Sec.~\ref{sec:r2sugra} we introduce $R^2$ inflation in SUGRA proposed in \cite{Ketov:2010qz} and further analyzed in \cite{Ketov:2012se}.
Then in Sec.~\ref{sec:flat} we incorporate a SUSY flat direction $\phi$ to analyze non-Gaussianity produced in this model. 
Sec.~\ref{sec:conclusion} is devoted to conclusion.
In the appendix we describe scalaron interactions with other fields and
calculate its decay widths as
an equivalent way to calculate gravitational particle production rate.

\section{$R^2$ inflation in SUGRA}\label{sec:r2sugra}
Here first we briefly introduce higher curvature inflation model
proposed by Ketov \cite{Ketov:2010eg} and Ketov and Starobinsky
\cite{Ketov:2010qz} in SUGRA.  The theory is defined by the action
\be
S &=&  \kappa^{-1}\int d^4xd^2\theta {\cal E}F({\cal R}) + {\rm h.c.}, \label{act}
\ee
where $F(\calR)$ is a function of the scalar curvature superfield
\be
{\cal R} &=& -\frac{\kappa}{3}B^* -\theta\left(\sigma^a\bar{\sigma}^b\psi_{ab} -i\sigma^a\bar{\psi}_a\frac{\kappa}{3}B^* + i\psi_a b^a \right) \non\\
&&+\theta\theta\left(\frac{\kappa}{3}R-\frac{i\kappa}{2}\epsilon^{abcd}R_{abcd}-\frac{4\kappa^3}{9}BB^* +\cdots \right),
\ee
and
$\cal E$ is the chiral superspace density in a Wess-Zumino type gauge
\be
{\cal E} &=& e(x) \left[ 1-2i\theta\sigma_a\bar{\psi}^a(x)+\theta\theta
\kappa^2B(x) \right]. \label{ep}
\ee
Here, $e=\sqrt{-g}$, $\psi^a$ is the gravitino, $B$ is an auxiliary
scalar field, and  $\kappa$ is the reciprocal of the reduced Planck scale
$\mg=\kappa^{-1}=m_{Pl}/\sqrt{8\pi}=2.4\times 10^{18}$ GeV.

Ignoring fermions, we find
\be
F({\cal R}) &=&F\left(-\frac{\kappa}{3}B^* \right) 
+F'\left(-\frac{\kappa}{3}B^* \right) \left(\frac{\kappa}{3}R
-\frac{4\kappa^3}{9}BB^* \right)\theta\theta,\\
S&=&\int d^4x \, e\left[ -3X F\left(X^* \right) +F'\left(X^* \right) \left(\frac13R-4XX^*\right)\right] + {\rm h.c.},\\
X&\equiv &-\frac{\kappa}{3}B.
\ee
Ignoring a pseudo-scalar partner (axion) of the scalaron, namely, taking
$X=X^*$ and $F=F^*$, the Lagrangian of the gravity sector reads,
\be
{\cal L} &=& -6XF(X) +2\left(\frac{1}{3} {R}-4X^2\right)F'(X), \label{LG}
\ee
together with the constraint which is nothing but the equation of motion
(EOM) for the auxiliary field
\be
0&=&3F(X)+11XF'(X)-\left(\frac{1}{3} {R}-4X^2\right)F''(X). \label{XEOM}
\ee
The choice of Ketov and Starobinsky is
\be
F({\cal R}) =\frac12 f_1{\cal R}+\frac12f_2{\cal R}^2 +\frac16 f_3{\cal R}^3,
\ee
which leads the Lagrangian
\be
{\cal L} &=& \frac13 f_1{R}+\frac23 f_2{R}X +\left(\frac13 f_3R -7f_1\right)X^2-11f_2 X^3-5f_3X^4 ,
\ee
and the constraint
\be
0&=&X^3 +\frac{33f_2}{20f_3}X^2-\frac{1}{30}(R-R_0)X
-\frac{f_2}{30f_3}R,\label{eq:field-X} 
\ee
with
\be
R_0&\equiv& \frac{21f_1}{f_3},
\ee
where $f_1$, $f_2$, and $f_3$ are positive constants.  Note that we use the metric signature $(-,+,+,+)$ so that
our sign convention of ${\cal R}$ and $R=12H^2+6\dot{H}$ is different from that of Ketov and Starobinsky.

Since the gravitational constant must be positive for 
stability of the system, $F'(X)>0$ and thus $f_1f_3>f_2^2$.
The absence of ghost and tachyonic degrees of freedom requires $f_1>0$ and $f_3>0$.
The sub-Planckian curvature during inflation requires $f_3 \gg 1$.
The sub-Planckian scalaron mass after inflation requires $f_2^2 \gg f_1$ 
in order to avoid large quantum gravity loop corrections.
In summary, \cite{Ketov:2010qz}
\be
f_3 \gg 1, \quad 
f_1f_3 > f_2^2 \gg f_1 >0. \label{cond}
\ee

In terms of the reduced Planck scale $\mg$, the scalaron mass during
inflation, $M$, and that in the reheating stage around the potential
minimum, $m$, they can be expressed as
\be
 f_1=\frac{3}{2}\mg^2,~~~f_2=\sqrt{\frac{63}{8}}\frac{\mg^2}{m},~~~
{\rm and}~~~f_3=\frac{15\mg^2}{M^2},
\ee
respectively.
Then the conditions (\ref{cond}) read
\be
m>\sqrt{\frac{7}{20}}M,~~~M\ll \mg,~~~m\ll \mg,
\ee
respectively \cite{Ketov:2012se}.

In the high curvature regime $R \gg R_0 = 21f_1/f_3=21M^2/10$,
the Lagrangian reads
\be
 \calL &=&
 \frac{1}{3}f_1R+\frac{1}{180}f_3 R^2+\frac{\sqrt{30}}{100}f_2R^{3/2}
 \nonumber \\
 &=&\frac{\mg^2}{2}\lmk R+\frac{R^2}{6M^2}+
\frac{3\sqrt{105}}{100}\frac{R^{3/2}}{m}\rmk. \label{effR}
\ee
The dynamics of inflation realized in this Lagrangian has been studied
by Ketov and Tsujikawa \cite{Ketov:2012se} in detail.  They solved the EOM
 for the Hubble parameter of the Friedmann Universe, $H(t)$, and
expressed the number of $e$-folds, $N$, acquired during inflation after the
epoch $H(t)=H$ in terms of the dimensionless quantities
\be
 \alpha\equiv \frac{M^2}{mH}~~{\rm and }~~\beta\equiv \frac{M^2}{H^2},
\ee
as 
\begin{eqnarray}
N&=& \frac{1}{126\alpha^2}\lkk 3\alpha\lmk 80\sqrt{35}-21\alpha-
\sqrt{7(63\alpha^2+16000\beta/3)}\rmk-\frac{4000\beta}{3}(8\ln 2+3\ln 5)
\right.\nonumber \\
&&\left. 
+\frac{8000\beta}{3}\ln\lmk\frac{\sqrt{7}(63\alpha^2+800\beta)
+21\alpha\sqrt{63\alpha^2+1600\beta}}{21\alpha+2\sqrt{35}\beta}\rmk\rkk.
\end{eqnarray}
The square amplitude of curvature perturbation 
is given by
\begin{eqnarray}
 \calP_{\zetar}(N) \cong \frac{1250}{3\pi^2}\lmk\frac{M}{\mg}\rmk^2
\left(3\sqrt{35}\alpha+\frac{100\beta}{3}\right)^{-2} . 
\end{eqnarray}
We assume this curvature perturbation generated by inflation is
responsible for the  fraction
$\lambda^2$ of the total amplitude, $\calP_{obs}\cong2.4\times 10^{-9}$,  
determined by COBE and WMAP \cite{Hinshaw:2012fq} 
on the pivot scale $k_*=0.002$Mpc$^{-1}$,
namely,
\begin{eqnarray} 
\calP_{\zetar}(N_*)= \lambda^2\calP_{obs}= 2.4\times 10^{-9}\lambda^2,
\end{eqnarray}
where $N_*$ is the number of $e$-folds of inflation after the
comoving pivot scale left the Hubble radius.

 The scalar spectral index $n_s$ and the tensor-to-scalar ratio
$r$ can also be expressed by $\alpha$ and $\beta$ as
\be
 n_s=1-\frac{3\sqrt{35}}{100}\alpha-\frac{2\beta}{3},
~~{\rm and}~~r=\frac{1}{2500}(9\sqrt{35}\alpha+100\beta)^2,
\ee
for the case $\lambda^2=1$ \cite{Ketov:2012se}.

As is seen in (\ref{effR}), the original $R^2$-inflation is obtained in
the small $\alpha$ limit, when $N$ approaches $3/\beta-1/2$
or $\beta=6/(2N+1)$.
Then the COBE-WMAP normalization yields
\be
  M=\frac{7.54\times 10^{-4}}{N_*+1/2}\lambda\mg=1.36\times
  10^{-5}\lambda\nus^{-1}\mg,
~~~\nus\equiv \frac{N_*+1/2}{55.5} \label{fixM}
\ee
which in turn determines the Hubble parameter as a function of $N$
as
\be
 H=\lmk\frac{N+1/2}{6}\rmk^{1/2}M.
  \label{HubN} 
\ee
Thus the Hubble parameter when the pivot scale left the Hubble radius
is given by
\be
H_*=4.1\times 10^{-5}\lambda\nus^{-1/2}\mg. \label{hubble-exit}
\ee

On the other hand, the mass of the scalaron in the reheating phase is 
determined as
\be
  m=\frac{M^2}{\alpha_* H_*}=4.47\times 10^{-6}\lambda\nus^{-3/2}\alpha_*^{-1}\mg,
\ee
so that the scalaron mass in the reheating phase is unrelated with
the CMB observation, and can be chosen arbitrarily under the condition
$m \ll \mg$.  Here $\alpha_*$ is the value of $\alpha$ at $H=H_*$.
Thus the decay rate of the scalaron (see Appendix A and set $M_s=m$ there),
\be
  \Gamma\cong \NS\GSK=\frac{\NS m^3}{192\pi\mg^2}, \label{totdecay}
\ee
where $\NS$ stands for the number of scalar decay modes due to scalar kinetic interaction terms, can be larger than
the cosmic expansion rate at the end of inflation, 
 $H_f\cong M/\sqrt{12}$, if
\be
m>0.133 \NS^{-1/3}\nus^{-1/3}\lambda^{1/3}\mg\equiv m_{rh}. \label{mrh}
\ee 
In such cases, the Universe is reheated efficiently soon after inflation
even if the inflaton-scalaron decays only through gravitational
interaction.  In what follows we fix $m=m_{rh}$ so that the Universe is
reheated rapidly after inflation.

The above inequality (\ref{mrh}) imposes an upper bound on 
$\alpha$ as 
\be
\alpha_* < 1.6\times 10^{-4}\lambda^{2/3}
\nus^{-7/6}(\NS/10^2)^{1/3}.
\ee
  Thus the rapid reheating after inflation
automatically guarantees the original $R^2$-inflation in the
supergravity context.  The scalar spectral index $\nsr$ and the
tensor-to-scalar ratio $r_{R2}$ from inflation are given by
\be 
\nsr=1-\frac{4}{2N+1}=0.964,~~~r_{R2}=\frac{48}{(2N+1)^2}=3.9\times 10^{-3},
\label{nsr2}
\ee
respectively, where we have quoted the numerical values at $N=55$.

Finally we comment on the thermal history after inflation in this model.
It has been known that finite amount of radiation is created 
after the onset of field oscillation 
even if
only perturbative decay with the rate (\ref{totdecay}) operates,
and its  energy density is given by 
$\rho_r(t)=\frac{6}{5}\Gamma H(t)\mg^2$
(see e.g. \cite{JY}).  Since $\Gamma$ is as high as $H_f$ here by assumption, the
radiation temperature just after inflation is comparable---in fact
only 20\% lower--- to the case all
the energy density is immediately converted to radiation at the end of 
inflation.  Thus even if the preheating may occur just after inflation as
analyzed by Ketov and Tsujikawa \cite{Ketov:2012se}, the
thermal history does not change significantly.  Whether preheating 
occurs or not in the presence of large $\Gamma$ deserves further study,
though.  The high temperature effect may modify the potential of the
flat direction to initiate field oscillation due to thermal effects,
which does not change $\delta\phi/\phi$ because both the homogeneous
mode $\phi$ and superhorizon long-wave fluctuations $\delta\phi$ 
would oscillate in the same
manner.  Furthermore,
in our model, the scalaron decays rapidly before these oscillations 
dissipate field amplitude appreciably.  So we may use the field values
at the end of inflation below.

Since the reheat temperature is quite high, the gravitino problem is a
problem \cite{gravitino}.  We should either adopt a model with large enough gravitino
mass without changing cosmic expansion history, or invoke non-standard
cosmology like thermal inflation \cite{thinf}.  In the former case, the 
pivot scale is pushed up to $N_*=60$ due to the earlier radiation
domination, 
while in the latter case it may be
shifted to a value as small as  $N_* \sim 50$.

\section{Modulated reheating through SUSY flat directions}\label{sec:flat}

We now consider behaviors of a SUSY flat direction field  in this
inflation model.  Although our favorite one is $H_uH_d$ flat direction
in close 
analogue to the Higgs condensation in the SM, other combinations such
as $LH_u$ or those involving three fields  would also work
with their appropriate choice of coupling constants, provided they do
not suffer from Q-ball instability before the scalaron decay.

Here we denote a generic flat direction by $\phi$ and study its
cosmological effects.\footnote{Note that a flat direction $\phi$ is not the inflaton, but causes curvature perturbations by modulated reheating. The $H_uH_d$ flat direction can be associated with it by $^tH_u\propto (0,\phi)$ and $^tH_d\propto (\phi,0)$.}  
By definition $\phi$ acquires a potential only
through SUSY breaking and possible nonrenormalizable terms in the
superpotential, $W \supset \frac{(\sqrt{2}\Phi)^n}{n\mx^{n-3}}$ with
$\Phi=\frac{\phi}{\sqrt{2}}e^{i\delta}$.  Since we are interested only in 
 the (position-dependent) amplitude of the flat direction, we suppress
possible CP violating A-terms to write the potential as
\be
  V[\phi]=\frac{1}{2}m_0^2\phi^2+\frac{\phi^{2(n-1)}}{\mx^{2(n-3)}},~~~
n\geq 4,
\ee
for $\phi \ll \mg$.  Here $m_0$ is the SUSY breaking mass assumed
to be $\order (1\rm{TeV})$ or so, and $\mx$ is a cutoff scale to 
characterize nonrenormalizable interactions.

The second term is  dominant for $\phi >
m_0^{\frac{1}{n-2}}\mx^{\frac{n-3}{n-2}}\equiv \phi_{NR}$ which we
assume is satisfied during inflation in the subsequent analysis.
For $n=4$ and 6 we find $\phi_{NR}=2.0\times 10^{-8}\mg$ and
$1.4\times 10^{-4}\mg$, respectively, for $m_0=1$TeV and $\mx=\mg$.
With no Hubble induced mass terms here during inflation, the initial value of
$\phi$ is randomly distributed, and it does not evolve significantly
during inflation if its effective mass is smaller than the Hubble
parameter, $V''[\phi] < H_f^2$.  This yields an upper bound on $\phi$
as 
\be
 \phi < \lkk\frac{H_f^2\mx^{2n-6}}{(2n-2)(2n-3)}\rkk^{\frac{1}{2n-4}}
\equiv \phi_{\max}. \label{phimaxn}
\ee
If it was not satisfied in the beginning, the field would  roll
down the potential until it is satisfied.
Using (\ref{fixM}) and (\ref{HubN}), we find 
\begin{eqnarray}
\phi_{\max}&=&8.5\times 10^{-4} \lambda^{1/2}
\nus^{-1/2}\lmk\frac{\mx}{\mg}\rmk^{1/2}\mg,~~~~(n=4), \nonumber\\
\phi_{\max}&=&2.5\times  10^{-2}\lambda^{1/4}
\nus^{-1/4}\lmk\frac{\mx}{\mg}\rmk^{3/4}\mg,~~~~(n=6),  \label{phimax}
\ee
respectively.

It is convenient to express solutions of the homogeneous mode, $\phi$,
and fluctuation around it, $\delta\phi$, as a function of the 
number of $e$-folds until the end of inflation, $N$, which is a 
decreasing function of cosmic time, using the Hubble parameter
(\ref{HubN}) \cite{Suyama:2007bg}.  The slow-roll EOM of $\phi$
reads
\be
 \frac{d\phi}{dN}=\frac{1}{3-H'/H}\frac{V_\phi}{H^2}
=\frac{6V_\phi}{(3N+1)M^2},\quad
H' = \frac{dH}{dN},
\ee
which is solved as
\be
 \phi(N)=\phis\lkk 1+\frac{8(n-1)(n-2)\phis^{2n-4}}{M^2\mx^{2n-6}}
\ln\lmk\frac{N_*+1/3}{N+1/3}\rmk\rkk^{\frac{-1}{2n-4}}.
\ee
Here $\phis$ is the initial value at $N=N_*$, which we take the
epoch of the wave number of our interest, namely, the pivot scale
$k_*$ left the Hubble radius during inflation, and a prime denotes
differentiation with respect to $N$.  In particular, at the end of
inflation corresponding to $N=0$, we find
\begin{eqnarray}
 \phi(0)&=&\phis\lkk 1+\frac{8(n-1)(n-2)\phis^{2n-4}}{M^2\mx^{2n-6}}
\ln\lmk 3N_*+1\rmk\rkk^{\frac{-1}{2n-4}}\!\!\!
\equiv\phif =
\phis(1+\Delta)^{\frac{-1}{2n-4}},\non\\ 
\Delta &\equiv& 
\frac{8(n-1)(n-2)\phis^{2n-4}}{M^2\mx^{2n-6}}
\ln\lmk 3N_*+1\rmk < \frac{n-2}{3(2n-3)}\ln{(3N_*+1)}\equiv \Delta_{\max}, \label{Delta}
\end{eqnarray}
where we have used (\ref{phimaxn}) and $H_f \cong M/\sqrt{12}$.
We find $\Delta_{\max}^{n=4}=0.682$ and $\Delta_{\max}^{n=6}=0.757$ for $N_*=55$, respectively.

Similarly, the super-horizon evolution of fluctuation can be obtained by
perturbatively solving the EOM 
\be
 \frac{d\delta\phi}{dN}=\frac{1}{3-H'/H}
 \lkk \frac{V_{\phi\phi}}{H^2}\delta\phi +\frac{V_{\phi\phi\phi}}{2H^2}
(\delta\phi)^2\rkk,
\ee
to yield
\begin{eqnarray}
\delta\phi(N) &=& \delta\phis\exp\lmk \int_{N_*}^N
\frac{1}{3-H'/H}\frac{V_{\phi\phi}}{H^2}dN\rmk \nonumber\\
&+&\frac{1}{2}(\delta\phis)^2\int_{N_*}^N
\frac{dN'}{3-H'/H}\frac{V_{\phi\phi\phi}}{H^2}
\exp\lmk  \int_{N_*}^{N'}\frac{1}{3-H'/H}\frac{V_{\phi\phi}}{H^2}
dN'' \rmk\\
&&~~~\times \exp\lmk \int_{N_*}^N
\frac{1}{3-H'/H}\frac{V_{\phi\phi}}{H^2}dN\rmk. \non
\end{eqnarray}
Here the initial condition is given by $\delta\phis =H_*/(2\pi)$. 
At the end of inflation,
we find
\begin{eqnarray}
\delta\phif &=&
 \delta\phis\lmk\frac{\phif}{\phis}\rmk^{2n-3}+\frac{1}{2}
\Theta(\delta\phis)^2, \label{deltaphif}
\end{eqnarray}
where $\Theta$ is given by
\begin{eqnarray}
\Theta &=& -\frac{2n-3}{\phis^{2n-3}}\lmk \phis^{2n-4}-
\phif^{2n-4}\rmk\lmk\frac{\phif}{\phis}\rmk^{2n-3}\nonumber\\
&=&-\frac{(2n-3)\Delta}{(1+\Delta)^{2+\frac{1}{n-2}}\phif}\equiv
-\frac{\Upsilon}{\phif}.
\end{eqnarray}
To calculate the power spectrum of curvature perturbation generated
by modulated reheating, only the first term in (\ref{deltaphif}) is
important, whereas the second term is also necessary 
in principle to calculate non-Gaussianity.

Since the Universe is reheated rapidly 
 after inflation, we can estimate the
amplitude of curvature perturbation produced by the
modulated reheating, $\zetam$, and its non-Gaussianity using
$\phif$ and $\delta\phif$.
For this purpose it is convenient to use the scalaron picture
and use the formulas given in \cite{Suyama:2007bg}.

First $\zetam$ is given by
\be
 \zetam=xQ'(x)\frac{\delta\Gamma}{\Gamma},
\ee
where $Q(x)$ is defined by
\be
  Q(x)=Q(\Gamma/H)=\frac{1}{4}\ln \lkk
 \int_0^\infty dN'\frac{\Gamma}{H(N')}e^{4N'}\frac{\rho_\varphi(N')}
{\rho_{\varphi f}}\rkk.
\ee
Here $\rho_\varphi$ is the effective energy density of the
scalaron-inflaton and  the subscript $f$ denotes the value
at the end of inflation \cite{Suyama:2007bg}.  Our choice of the mass parameter corresponds
to $x=1$, where we find $Q'(1)=-0.0890$ and $Q''(1)=0.126$.
Here the total decay rate $\Gamma$ is given by
$
\Gamma \cong {\cal N}_{S}\GSK
$
(${\cal N}_{S} ={\cal N}_{\sigma}+{\cal N}_{\phi}+$ all other light scalars) and the space-dependent part of the rate is expressed as
\begin{eqnarray}
\delta\Gamma &=&\Gamma_\phi\delta\phi
= {\cal N}_{\sigma}\Gamma_{SKM\phi}\delta\phi+{\cal N}_{\psi}\Gamma_{FM\phi}\delta\phi \non\\ 
&=&\frac{({\cal N}_{\sigma} + {\cal N}_{\psi})y^2m\phi\delta\phi}{24\pi\mg^2} 
\equiv \frac{\NFL y^2m\phi\delta\phi}{24\pi\mg^2},
\end{eqnarray}
where we have put $h=y$ and replaced $M_s$ by $m$ in the formulas
(\ref{scalarwidth}) and (\ref{fermionwidth}).
Thus $\zetam$ is given by
\begin{eqnarray}
 -\zetam&=&0.712\frac{\NFL y^2\phif\delta\phif}{\NS^{1/3}m^2}
=2.33\times
10^3\mu\nus^{2/3}\lambda^{-2/3}\frac{\phif\delta\phif}{\mg^2}, \non\\
\mu&\equiv&\lmk\frac{y}{3}\rmk^2\lmk\frac{\NFL}{30}\rmk
\lmk\frac{\NS}{10^2}\rmk^{-1/3}.
\end{eqnarray}
Here $\mu$ is a numerical factor normalized by  typical values
of the number of decay modes and Yukawa coupling.  Note that
the number of decay modes for a pair of a single species of quark and
antiquark, say $t$ and $\overline{t}$, is $\calN_\psi=12$
and supersymmetry doubles it by the same contribution of $\GSKM$. 
We also note that Yukawa coupling is typically large, which
might even saturate perturbative bound $y^2/(4\pi)\sim 1$ at high
energy scale of our concern \cite{Barger:1992ac}.
Using (\ref{Delta}), (\ref{deltaphif}) and (\ref{hubble-exit}), we find 
\be
 -\zetam=1.53\times 10^{-2}\mu\nus^{1/6}\lambda^{1/3}(1+\Delta)^{-\frac{n-1}{n-2}}
\frac{\phis}{\mg}. \label{zetam}
\ee
It should contribute to the fraction $1-\lambda^2$ of the observed
power spectrum by assumption, so $\calP_{\zetam}=(1-\lambda^2)\calP_{obs}$.
Then $\phis/\mg$ can be expressed as
\be
\frac{\phis}{\mg}=3.2\times 10^{-3} \mu^{-1}\nus^{-1/6}\lambda^{-1/3}
(1-\lambda^2)^{1/2}(1+\Delta)^{\frac{n-1}{n-2}}.  \label{phis}
\ee
The spectral index, $\nsm$, of perturbation produced by modulated reheating
is given by
\be
 \nsm\cong 1-\frac{2}{2N+1}=0.982,  \label{nsm}
\ee
where the last equality is for the case $N=55$.

Next we turn to the non-Gaussianity.  It is known that modulated
reheating produces an appreciable amount of non-Gaussianity of local
type \cite{Zaldarriaga:2003my,Suyama:2007bg}, 
and the nonlinearity parameter $\fnl$ of $\zetam$, $\fnlm$,
is given by
\begin{eqnarray}
\fnlm=\frac{5Q''(x)}{6Q'^2(x)}+\frac{5}{6xQ'(x)}\lmk
\frac{\Gamma\Gamma_{\phi\phi}}{\Gamma^2_\phi}+\frac{\Gamma\Theta}{\Gamma_\phi}\rmk
= 13.2-1.17(1-\Upsilon)\frac{\NS m^2}{\NFL y^2\phif^2}
\end{eqnarray}
for $x=1$.
The full nonlinearity parameter is given by
\begin{eqnarray}
\fnl&=&\fnlm\frac{\calP_{\zetam}^2}{\calP_{obs}^2}
=\fnlm (1-\lambda^2)^2 \nonumber\\
&=&13.2 (1-\lambda^2)^2+35.0(\Upsilon-1)(1+\Delta)^{-\frac{2n-3}{n-2}}\mu\nus^{-1/3} 
\lambda^{4/3}(1-\lambda^2), \label{fnlformula}
\end{eqnarray}
where we have used (\ref{mrh}), (\ref{Delta}) and (\ref{zetam}).

From (\ref{Delta}), (\ref{phis}) and (\ref{fixM}) $\Delta$ satisfies
\be
\Delta=1.38\times
10^2 \frac{\ln{(3N_*+1)}}{\ln{166}}\nus^{4/3}\mu^{-4}(1+\Delta)^6\lambda^{-10/3}(1-\lambda^2)^2
\lmk\frac{\mg}{\mx}\rmk^2 \label{delta4}
\ee
for $n=4$, and
\be
\Delta=4.78\times
10^{-8}  \frac{\ln{(3N_*+1)}}{\ln{166}}\nus^{2/3}\mu^{-8}(1+\Delta)^{10}\lambda^{-14/3}(1-\lambda^2)^4
\lmk\frac{\mg}{\mx}\rmk^6 \label{delta6}
\ee
for $n=6$.  Here note that we are not allowed to take $\lambda=0$
exactly, because we have first fixed $M$ from (\ref{fixM}) and this procedure 
is invalid there.  Since this problem arises if and only if we take
$\lambda$ exactly equal to zero from the beginning, our formula for $\fnl$
and $n_s$ below are applicable even in the limit $\lambda$ is small and
approaches to zero.

For definiteness, let us take $\nus=1$ (i.e. $N_*=55$) and $\mu=1$ below.  Then we can
express $\fnl$ as a function of $\lambda$ and $\Delta$ for each $n$ as
\be
 \fnl=13.2(1-\lambda^2)^2+35.0K(n,\Delta)\lambda^{4/3}(1-\lambda^2),
\ee
where 
\be
 K(n,\Delta)\equiv (\Upsilon-1)(1+\Delta)^{-\frac{2n-3}{n-2}}
=(2n-3)\Delta(1+\Delta)^{-\frac{4n-6}{n-2}}-
(1+\Delta)^{-\frac{2n-3}{n-2}}.
\ee
We can also determine the cut off scale from (\ref{delta4}) and
(\ref{delta6}) as
\begin{eqnarray}
\mx&=&12\Delta^{-1/2}(1+\Delta)^3\lambda^{-5/3}(1-\lambda^2)\mg,~~~~~~~~~~~~~~~~~~~(n=4), \\
\mx&=&6.0\times 10^{-2}\Delta^{-1/6}
(1+\Delta)^{5/3}\lambda^{-7/9}(1-\lambda^2)^{2/3}\mg,~~~(n=6),
\end{eqnarray}
respectively.

Now let us evaluate $K(n,\Delta)$ for $n=4$ and $n=6$.  We find
\be
 -1 \le K(4,\Delta) \le  -1.93\times 10^{-2},
\ee
which is monotonically increasing as $\Delta$, and the inequalities are saturated at $\Delta=0$ and $\Delta_{\max}=0.682$,
respectively.  For $n=6$ we find
\be
 -1 \le K(6,\Delta) \le 0.327,
\ee
where the minimum is realized again at $\Delta=0$ and the
maximum at $\Delta=0.451$.

The case $\Delta \longrightarrow 0$ corresponds to the limit
$\mx$ is large and nonrenormalizable interaction is absent.
Then, independent of $n$ we find
\be
\fnl=13.2(1-\lambda^2)^2-35.0\lambda^{4/3}(1-\lambda^2),
\ee
which takes maximum $\fnl=13.2$ in the limit $\lambda\longrightarrow 0$,
and minimum $\fnl=-7.9$ at $\lambda=0.749$.  The former corresponds to the case
all the observed fluctuations are due to the modulated reheating 
with the spectral index $n_s=0.982$, while at $\lambda=0.749$ we find
$n_s=0.972$ for $\nus=1$, because the  spectral index
of total curvature perturbation, $n_s$ is given by 
\be
 n_s=\lambda^2 \nsr + (1-\lambda^2)\nsm=1-\frac{2\lambda^2+2}{111\nus}, \label{nsformula}
\ee
in general.

On the other hand, with $K(4,0.682)=-1.93\times 10^{-2}$,
one finds 
\be
\fnl=13.2(1-\lambda^2)^2-0.675\lambda^{4/3}(1-\lambda^2).
\ee
In this case $\fnl$ decreases monotonically from 13.2 to $-8.32\times 10^{-3}$ as we increase
$\lambda$ from 0 to 0.988. Then $\fnl$ increases to 0 as $\lambda\to 1$.

For the case $K(6,0.451)=0.327$ we have
\be
\fnl=13.2(1-\lambda^2)^2+11.5\lambda^{4/3}(1-\lambda^2),
\ee
which takes maximum $\fnl=13.5$ at $\lambda = 0.148$
and then decreases to $0$ as $\lambda$ is increased.

Our results suggest that as far as $\fnl$ parameter is concerned,
the nonrenormalizable term in the potential of the
flat direction does not have observationally significant effects,
which means that our result is robust in this respect.

If we take smaller values for $\mu$, then the second term in (\ref{fnlformula}) contributes less and the overall $\fnl$ can become larger for $n=4$ and smaller for $n=6$ than the demonstrated case above.

\section{Conclusion}\label{sec:conclusion}

In the present paper we have reconsidered cosmic history after
$R^2$-inflation in SUGRA in which reheating proceeds through 
gravitational particle production of conformally noninvariant fields.
We have argued that conformal invariance is broken through a
nonvanishing
expectation value of a SUSY flat direction which gives a position
dependent mass for those fields coupled to it due to its long-wave
quantum fluctuations acquired during inflation.  As a result
such fluctuations induce modulated reheating.

We have shown that with reasonable values of the model parameters,
on condition that the universe is reheated rapidly, curvature
perturbation produced by modulated reheating may contribute to the
observed fluctuation significantly and yields a local non-Gaussianity
with $\fnl \sim \pm 10$.

The intriguing feature of our model is that the observable parameters
$n_s,~r,~\fnl$ are mutually correlated with each other through the
ratio of the two contributions $\lambda^2$, although the exact one-to-one
correspondence is difficult due to the uncertainties in the pivot
scale $\nus$ and the particle-physics parameter $\mu$.
For example, from (\ref{nsformula}) and 
\be
r=r_{R2}\lambda^2= \frac{48\lambda^2}{(2N_*+1)^2}=\frac{48\lambda^2}{111^2\nus^2}  , \label{rformula}
\ee
one can obtain $\lambda$ and $\nus$ (or $N_*$), which can be compared with $\fnl$.
In future, information on the thermal history carried by $\nus$ can
also be tested by space based laser interferometer such as DECIGO 
\cite{Seto:2001qf,suwa}, and can be compared with our result.

\subsection{Note added after Planck 2013 results}\label{subsec:planck}
After we submitted the original version of 
this article to the arXiv, the Planck collaboration 
released their data. It gives \cite{Ade:2013rta,Ade:2013ydc}
\be
n_s = 0.9653 \pm 0.0069 \ ,\quad
r < 0.13~~(95\%~{\rm CL}),\quad
f_{NL} =2.7 \pm 5.8 ~~(68\%~{\rm CL}),
\ee 
where we have quoted values from Planck combined with WMAP polarization and lensing data sets for $n_s$ and $r$.
Their results are basically in good agreement with the predictions of
our model.

To be precise,  
the lower value of $n_s$, which favors larger $\lambda$, 
does not yield any sizable non-Gaussianity with our model. 
On the other hand, the higher allowed value like $n_s=0.967$ yields 
\be
r = 3.2\times 10^{-3} ,\quad
f_{NL} = \{ -4.8,\  0.28,\  2.1 \}
\ee
for $\{ K(n,0_+), K(4,0.68), K(6,0.45) \}$, $\lambda = 0.91$, $\mu =1$ 
and $\nu_*=1$.
Thus our model covers lower allowed range of the observed $f_{NL}$,
namely, between $-0.1\sigma$ and $-1.3\sigma$ off the mean value.

From (\ref{nsformula}) one can see that smaller
 $\nu_*$ (or $N_*$) allows smaller $\lambda$ resulting in higher values
 for $|\fnl|$. The thermal inflation scenario \cite{thinf} 
after $R^2$ inflation may yield such values. 

Finally, let us comment on the SM Higgs inflation with nonminimal
coupling to gravity \cite{Bezrukov:2007ep}, which is also preferred by
the Planck data \cite{Ade:2013rta}. As was emphasized in
\cite{Bezrukov:2011gp}, predictions of $n_s$ and $r$ from the Higgs
inflation and $R^2$ inflation are almost degenerate. Only the slight
difference comes from the reheating stage. In fact, the SUSY flat
direction alters reheating significantly from that after the original
$R^2$ inflation, thereby producing non-negligible $f_{NL}$ as 
opposed to the Higgs inflation. In our scenario the reheating 
temperature can be as high as $10^{14}$ GeV and spatially modulated.

\appendix
\section{Scalaron interaction with matter fields}\label{appendix}

In generic $f(R)$-type gravitational theory,
\be
S =   \frac{1}{2\kappa^2}\int f(R) \sqrt{-g}d^4x +S_m,
\ee
the extra scalar degree of freedom, $\varphi$, of gravity is extracted by
\be
 \varphi\equiv \sqrt{\frac{3}{2}}\mg\ln|f'(R)|,
\ee
and its mass $M_s$ is given by $M_s=1/\sqrt{3f''(R)}$.
So in the original $R^2$-inflation model with $f(R)=R+R^2/(6M^2)$, the 
scalaron mass is constant and given by $M$ which is fixed by the
observed
amplitude of curvature fluctuations.  The supergravity model
discussed in this paper has a more complicated action and the
scalaron, which is in fact a complex field and we are focusing on
its real part throughout \cite{Ketov:2012se}, has different masses
at different regimes.

Separating this scalaron degree of freedom through appropriate conformal
rescaling, the rest of the gravitational action is given by the familiar
Einstein term only.  The action for a massive scalar field $\sigma$,
after the conformal rescaling $\sigma \longrightarrow 
e^{-\frac{\kappa\varphi}{\sqrt{6}}}\sigma$ reads
\be\label{eq:scalar-einstein}
{\cal L}_{\rm scalar} = -\frac12 
\partial_{\mu}\sigma\partial^{\mu}\sigma 
-\frac{\kappa\sigma}{\sqrt6} 
\partial_{\mu}\sigma\partial^{\mu}\varphi 
-\frac{\kappa^2 \sigma^2}{12}\partial_{\mu}\varphi\partial^{\mu}\varphi \non\\
- \frac{m_{\sigma}^2}{2} e^{-\frac{2}{\sqrt6}
\kappa\varphi}\sigma^2. 
\ee
Similarly, the action for a canonical fermion is given by
\be\label{eq:fermion-einstein}
{\cal L}_{\rm fermion} = - \bar\psi \Slash{D} \psi - e^{-\frac{1}{\sqrt6}\kappa\varphi}m_{\psi}\bar\psi\psi, 
\ee
after rescaling $\psi \longrightarrow 
e^{-\frac{\sqrt{6}\kappa\varphi}{4}}\psi$.
The spinor covariant derivative is defined, for a U(1) gauge field $A_{\mu}$, as
\be
\Slash{D}\psi &\equiv& e^{\mu}{}_{\alpha}\gamma^{\alpha}(\partial_{\mu}+\Gamma_{\mu}-igA_{\mu})\psi,\\
\Gamma_{\mu} &\equiv& \frac18[\gamma^{\alpha},\gamma^{\beta}]e^{\lambda}{}_{\alpha}\nabla_{\mu}e_{\lambda\beta},
\ee
where $e^{\mu}{}_{\alpha}$ is a tetrad (vierbein) field, rescaled as 
$e^{\mu}{}_{\alpha}\longrightarrow 
e^{-\frac{\kappa\varphi}{\sqrt{6}}}e^{\mu}{}_{\alpha} $, and $\Gamma_{\mu}$ is a spin connection for the Dirac spinor \cite{Watanabe:2010vy}.
For a U(1) gauge field ${A}_{\mu}$ and a charged massless complex scalar 
$\Phi \equiv {\chi}e^{ig\zeta}/\sqrt2$, the transformed action is given by
\be\label{eq:vector-einstein}
{\cal L}_{\rm vector} = -\frac14 F^{\mu\nu}F_{\mu\nu} - \frac12 g^{\mu\nu}\partial_{\mu}\chi\partial_{\nu}\chi
-\frac{\kappa\chi}{\sqrt6} g^{\mu\nu}\partial_{\mu}\chi\partial_{\nu}\varphi -\frac{\kappa^2 \chi^2}{12}g^{\mu\nu}\partial_{\mu}\varphi\partial_{\nu}\varphi \non\\
-\frac{g^2}{2}\chi^2g^{\mu\nu}V_{\mu}V_{\nu},
\ee
where $F_{\mu\nu}\equiv\partial_{\mu}V_{\nu}-\partial_{\nu}V_{\mu}
=\partial_{\mu}A_{\nu}-\partial_{\nu}A_{\mu}$, a gauge invariant field 
$V_{\mu}\equiv A_{\mu} +\partial_{\mu} \zeta$ has been defined, 
and rescaled as $\chi \longrightarrow e^{-\frac{\kappa\varphi}{\sqrt{6}}}{\chi}$.
If the gauge symmetry has been broken explicitly or {\it spontaneously
before rescaling}, one could consider the gauge-breaking mass term
\be\label{eq:gauge-breaking-mass}
{\cal L}_{\rm g.b.m.} = -\frac{m_V^2}{2}e^{-\frac{2\kappa\varphi}{\sqrt6}}V^2.
\ee

The scalaron $\varphi$ can decay into the scalar $\sigma$, fermion $\psi$, and massive vector $V_{\mu}$  via trilinear interactions:
\be
{\cal L}_{\rm 3leg}&=&-\frac{1}{\sqrt6 M_{\rm Pl}}\sigma\partial^{\mu}\sigma\partial_{\mu}\varphi + \frac{m_{\sigma}^2}{\sqrt6 M_{\rm Pl}} \varphi\sigma^2 
+ \frac{m_{\psi}^2}{\sqrt6 M_{\rm Pl}} \varphi\bar\psi\psi 
+ \frac{m_{V}^2}{\sqrt6 M_{\rm Pl}} \varphi V^2 \non\\ 
&=& \frac{1}{\sqrt6 M_{\rm Pl}}\varphi \partial^{\mu}\sigma\partial_{\mu}\sigma + \frac{2m_{\sigma}^2}{\sqrt6 M_{\rm Pl}} \varphi\sigma^2 
+ \frac{m_{\psi}^2}{\sqrt6 M_{\rm Pl}} \varphi\bar\psi\psi+ \frac{m_{V}^2}{\sqrt6 M_{\rm Pl}} \varphi V^2,
\ee
where we have integrated by parts, used $\delta (\sqrt{-g}{\cal L}_{\rm scalar})/\delta \sigma=0$, and left only linear terms in $\varphi$ from Eqs.~(\ref{eq:scalar-einstein}), (\ref{eq:fermion-einstein}) and (\ref{eq:gauge-breaking-mass}). 

The decay rates of the scalaron are then given by \cite{Gorbunov:2010bn, Watanabe:2006ku, Watanabe:2010vy}
\be
\Gamma(\varphi\to\sigma\sigma) &=& \frac{\calN_{\sigma}(M_s^2+2m_{\sigma}^2)^2}{192\pi M_{\rm Pl}^2 M_s}\left(1-\frac{4m_{\sigma}^2}{M_s^2}\right)^{1/2},\\
\Gamma(\varphi\to\bar\psi\psi) &=& \frac{\calN_{\psi}m_{\psi}^2 M_s}{48\pi M_{\rm Pl}^2}\left(1-\frac{4m_{\psi}^2}{M_s^2}\right)^{3/2},\\
\Gamma(\varphi\to VV) &=& \frac{\calN_{V}m_{V}^4}{48\pi M_{\rm Pl}^2M_s}\left(1-\frac{4m_{V}^2}{M_s^2}\right)^{1/2}
\lmk 1-\frac{4m_{V}^2}{M_s^2}+\frac{12m_{V}^4}{M_s^4} \rmk,
\ee
where $\calN_{\sigma}$, $\calN_{\psi}$ and $\calN_V$ are the number of
modes for each field.
Note that we have added gauge-breaking mass
term~(\ref{eq:gauge-breaking-mass}) to the vector
action~(\ref{eq:vector-einstein}); otherwise 
the decay channel to vector modes do not show up classically.

We are interested in the case masses of the decay products are given by
a spatially-dependent expectation value of a scalar field $\phi$, which
is
either the Higgs field in the SM or a flat direction field in SUSY, as
$m_\sigma=h\phi$, $m_\psi=y\phi$, and $m_V=g\phi$ with $h,~y,$ and $g$
representing, scalar, Yukawa, and gauge couplings, respectively.
 Hereafter we consider the case masses of
the decay products are much smaller than the parent particle, so that
the phase space suppression factors in the above decay widths are
negligible. Then we can classify each mode according to
the dependence on $\phi$ as follows.
\begin{eqnarray}
\Gamma(\varphi\to\sigma\sigma) &=& \frac{\calN_\sigma M_s^3}{192\pi\mg^2}
+\frac{\calN_\sigma (h\phi)^2M_s}{48\pi\mg^2}+\frac{\calN_\sigma (h\phi)^4}{48\pi \mg^2 M_s}\non\\
&\equiv& \calN_\sigma (\GSK+\GSKM+\GSM), \label{scalarwidth}\\
\Gamma(\varphi\to\bar\psi\psi) &=& \frac{\calN_\psi (y\phi)^2M_s}{48\pi\mg^2}
\equiv \calN_\psi\GFM, \label{fermionwidth}\\
\Gamma(\varphi\to VV) &=& \frac{\calN_V(g\phi)^4}{48\pi\mg^2 M_s}\equiv 
\calN_V\Gamma_{VM}.
\end{eqnarray}
Here $\GSK$ from scalar kinetic term is independent of $\phi$ and
provides the main part of the decay width, while $\GSKM$ and
$\GFM$ give equal contribution if $h=y$ which is the case 
in SUSY models.  The former is the interference contribution between
scalar kinetic and mass terms, and the latter is due to the fermionic mass
term.
Finally $\GSM$ and $\Gamma_{VM}$, both from their respective mass terms,
 are higher order in the
mass ratio compared with $\GSKM$ and $\GFM$, so that their contribution
will be minor, and we do not incorporate them to the subsequent
analysis, which turns out to be a good approximation in our case with
a large scalaron mass.
Note that the equivalence of the above scalaron-decay approach with the
Bogoliubov transformation approach 
to calculate particle production rate
have been shown in \cite{Vilenkin:1985md,Arbuzova:2011fu} 
in the present context and in \cite{Braden:2010wd}
in a different context.

\vskip 2cm
\noindent
{\large\bf Acknowledgements}

We are grateful to K.\ Hamaguchi, T.\ Moroi, K.\
Nakayama, K.\ Kamada, T.\ Suyama, and S.\ Yokoyama for useful
communications.
This work was partially supported by 
 JSPS  Grant-in-Aid for Scientific Research
23340058 (J.Y.), and the Grant-in-Aid for
Scientific Research on Innovative Areas No. 21111006 (J.Y.).

\end{document}